\def\papertitle{Simulation-Based Inference for Plate Reverb System Identification}
\def\paperauthorA{Dylan Sechet}
\def\paperauthorB{Marc Evrard}
\def\paperauthorC{Matthieu Kowalski}

\documentclass[twoside,a4paper]{article}
\usepackage{etoolbox}

\usepackage[taskA]{dafx26challenge} 

\usepackage{amsmath,amssymb,amsfonts,amsthm}
\usepackage{siunitx}
\usepackage{euscript}
\usepackage[T1]{fontenc}
\usepackage[utf8]{inputenc}
\usepackage{ifpdf}
\usepackage[english]{babel}
\usepackage{caption}
\usepackage{subfig} 
\usepackage{color}
\usepackage{booktabs}
\usepackage{microtype}
\usepackage{needspace}
\usepackage{lipsum}

\input glyphtounicode
\ninept

\newcounter{numauth}
\newcounter{listcnt}
\newcommand\authcnt[1]{\ifdefined#1 \stepcounter{numauth} \fi}

\newcommand\addauth[1]{
  \ifdefined#1
  \stepcounter{listcnt}
  \ifnum \value{listcnt}<\value{numauth}
  \appto\authorslist{, #1}
  \else
  \appto\authorslist{~and~#1}
  \fi
\fi}
\authcnt{\paperauthorB}
\authcnt{\paperauthorC}
\authcnt{\paperauthorD}
\authcnt{\paperauthorE}
\authcnt{\paperauthorF}
\authcnt{\paperauthorG}
\authcnt{\paperauthorH}
\authcnt{\paperauthorI}
\authcnt{\paperauthorJ}
\def\authorslist{\paperauthorA}
\addauth{\paperauthorB}
\addauth{\paperauthorC}
\addauth{\paperauthorD}
\addauth{\paperauthorE}
\addauth{\paperauthorF}
\addauth{\paperauthorG}
\addauth{\paperauthorH}
\addauth{\paperauthorI}
\addauth{\paperauthorJ}

\usepackage{times}

\newif\ifpdf
\ifx\pdfoutput\relax
\else
\ifcase\pdfoutput
\pdffalse
\else
\pdftrue
\fi
\fi

\ifpdf 
\usepackage[pdftex,
  pdftitle={\papertitle},
  pdfauthor={\authorslist},
  pdfsubject={Proceedings of the 29th International Conference on Digital Audio Effects (DAFx26)},
  colorlinks=false, 
  bookmarksnumbered, 
  pdfstartview=XYZ 
]{hyperref}
\usepackage[pdftex]{graphicx}
\else 
\usepackage[dvips]{epsfig,graphicx}
\usepackage[dvips,
  pdftitle={\papertitle},
  pdfauthor={\authorslist},
  pdfsubject={Proceedings of the 29th International Conference on Digital Audio Effects (DAFx26)},
  colorlinks=false, 
  bookmarksnumbered, 
  pdfstartview=XYZ 
]{hyperref}
\fi
\usepackage[hypcap=true]{caption}
\title{\papertitle}

\usepackage[nameinlink,capitalize]{cleveref}

\newcommand{\alv}{\boldsymbol{\alpha}}
\newcommand{\bx}{\mathbf{x}}

\affiliation
{\paperauthorA\,\sthanks{Thanks to the predecessors for the templates}}
{\href{https://dafx26.mit.edu}{Dept. of Electrical Engineering and Computer Science} \\ Massachusetts Institute of Technology \\ Cambridge, USA\\
  {\tt \href{mailto:dafx2026@gmail.com}{dafx2026@gmail.com}}
}

\affiliation
{\paperauthorA\, \paperauthorB\, \paperauthorC\,}
{{Laboratoire Interdisciplinaire des Sciences du Numérique} \\
  {Université Paris-Saclay, Inria, CNRS, CentraleSupélec}\\
  {\tt {firstname.lastname@lisn.fr}}
}

\begin{document}
\ifpdf 
\DeclareGraphicsExtensions{.png,.jpg,.pdf}
\else  
\DeclareGraphicsExtensions{.eps}
\fi


\maketitle

\begin{abstract}
  We address Task A of the 1st DAFx Parameter Estimation Challenge, which aims to retrieve the physical parameters of a plate model from an impulse response.
  To do so, we use the Simulation-Based Inference (SBI) framework, in which we train a neural network to estimate a density over plate parameters given an impulse response, using a dataset generated by the simulator.
  Inference for a new impulse response then requires only a forward pass through the network, without involving the simulator.

  For each test observation, we fine-tune a specific network: additional simulation rounds are performed
  by sampling parameters from the current estimated distribution, simulating
  the corresponding impulse responses, and fine-tuning to produce the specialized network.
\end{abstract}

\section{Introduction}
\label{sec:intro}
Plate reverberation has a long history as an audio effect, from the electromechanical units of the 1960s such as the EMT 140 to their present-day digital emulations. Its physical behaviour is by now well understood: a plate reverberator can be described as a thin metal plate undergoing small transverse vibrations, governed by the damped Kirchhoff-Love equation, and a range of accurate digital models have been built on this basis, whether by direct numerical simulation of the plate equation~\cite{bilbaoDigitalPlateReverberation2007} or by modal decomposition~\cite{ducceschiPlateReverberationDevelopment}.

While the forward problem is well understood, the inverse problem is considerably harder, and recovering either material or modal parameters of the response from audio impulse responses alone remains a challenging task. These difficulties motivated the 1st DAFx Parameter Estimation Challenge~\cite{ducceschi1stDAFxParameter}, a data challenge for plate-reverb system identification using a common simulator and evaluation protocol. In this work, we tackle Task~A of the challenge: given the impulse response of a simulated plate, estimate its identifiable physical parameters.

Parameter estimation for audio effects has been approached from several directions, commonly characterized as white, grey or black-box modelling depending on how much knowledge of the target system is assumed.
Black-box neural networks can emulate plate and spring reverberation with high fidelity~\cite{ramirezModelingPlateSpring2020}, but expose no physical parameters.
Grey-box methods keep a known processing structure and fit its parameters by gradient descent, as in differentiable digital signal processing~\cite{engel2020ddsp}.
Such gradient-based fitting relies on a differentiable forward model, which the challenge simulator is not.

Simulation-Based Inference (SBI) removes that requirement: it treats the simulator as a black box queried only through samples, learning the inverse map and returning a full posterior over the physical parameters.

\section{Problem Setup}
\label{sec:setup}

The challenge's reference model~\cite{ducceschi1stDAFxParameter} is a thin rectangular plate of side lengths $L_x \times L_y$ undergoing small transverse vibrations, governed by the damped Kirchhoff-Love equation with simply-supported boundaries and simulated by modal synthesis. The observed data is the impulse response (IR): the transverse displacement recorded at a readout point $(x_o, y_o)$ following a Dirac impulse applied at a fixed input point $(0.335\,L_x,\, 0.467\,L_y)$. Each IR is a \SI{5}{\second} displacement signal sampled at \SI{44.1}{\kilo\hertz}, giving us 220\,500 samples.

The raw material parameters (density $\rho$, thickness $h$, Young's modulus $E$, tension per unit length $T_0$) affect the IR only through the three invariants $\mu := \rho h$, $D/\mu$ and $T_0/\mu$, with $D$ the flexural rigidity. Task~A therefore asks to estimate, from a single IR, the six identifiable parameters
\begin{equation}
  \alv := \{\mu,\; D/\mu,\; T_0/\mu,\; L_y,\; x_o,\; y_o\},
  \label{eq:params}
\end{equation}
all remaining parameters being fixed across simulations. In the following, $\alv \in [0,1]^6$ is normalized to the unit hypercube by the ranges of \cref{tab:ranges}. The IRs are provided unnormalized, as the absolute amplitude scale is inversely proportional to $\mu$ and would be discarded by peak normalization.

\begin{table}[ht]
  \centering
  \caption{Ranges of the six estimated parameters, used to normalize the NMSE~\eqref{eq:nmse}. The invariant ranges ($\mu$, $D/\mu$, $T_0/\mu$) are those induced by the raw-parameter ranges~\cite{ducceschi1stDAFxParameter}; $L_x = \SI{1}{\meter}$ is fixed.}
  \resizebox{\columnwidth}{!}{%
    \begin{tabular}{llcc}
      \toprule
      Parameter & Symbol & Range & Unit \\
      \midrule
      Areal mass density    & $\mu$     & $[2.43,\ 106.15]$            & \si{\kilo\gram\per\meter\squared} \\
      Bending-to-mass ratio & $D/\mu$   & $[0.28,\ 201.2]$            & \si{\meter\tothe{4}\per\second\squared} \\
      Tension-to-mass ratio & $T_0/\mu$ & $[9.4{\times}10^{-5},\ 411.5]$ & \si{\meter\squared\per\second\squared} \\
      Plate length ($y$)    & $L_y$     & $[1.1,\ 4.0]$               & \si{\meter} \\
      Readout position ($x$) & $x_o$    & $[0.51,\ 1.0]$              & frac.\ $L_x$ \\
      Readout position ($y$) & $y_o$    & $[0.51,\ 1.0]$              & frac.\ $L_y$ \\
      \bottomrule
    \end{tabular}%
  }

  \label{tab:ranges}
\end{table}

Estimates are scored by the challenge's range-normalized mean squared error, averaged over the six parameters:
\begin{equation}
  \mathrm{NMSE} = \frac{1}{6} \sum_{i=1}^{6}
  \frac{\bigl(\alpha_i^{\text{est}} - \alpha_i^{\text{ref}}\bigr)^2}
  {\bigl(\alpha_i^{\max} - \alpha_i^{\min}\bigr)^2}.
  \label{eq:nmse}
\end{equation}
A constant predictor taking the midpoint of each range achieves a $\mathrm{NMSE}$ of roughly $83\times10^{-3}$, which serves as a random-guess reference~\cite{ducceschi1stDAFxParameter}.

\section{Simulation-Based Inference}
\label{sec:sbi}

Let $\bx \in \mathbb{R}^T$ be an observed impulse response of length~$T$.
The plate simulator defines a forward model $\bx = f(\alv)$.

We wish to evaluate the likelihood $p(\bx \mid \alv)$, but doing so in closed form would be intractable.
The goal of Simulation-Based Inference is to learn the inverse map: given $\bx$, recover a distribution over the parameters $\alv$ that could have produced it.
Concretely, we seek a neural density estimator $q_\theta(\alv \mid \bx)$ that approximates the true posterior
\begin{equation}
  p(\alv \mid \bx) \propto p(\bx \mid \alv)\, p(\alv)
  \label{eq:post}
\end{equation}
where $p(\alv)$ is the challenge's generating distribution: the raw parameters are sampled uniformly within their ranges, which induces a non-uniform distribution over the invariants $\mu$, $D/\mu$ and $T_0/\mu$.

Rather than evaluating $p(\bx \mid \alv)$ directly, $q_\theta$ is trained by maximum likelihood on a dataset $\{(\alv_n,\, \bx_n {:=} f(\alv_n))\}_{n=1}^{N}$ of simulator-generated pairs, minimizing:
\begin{equation}
  \mathcal{L}(\theta) := -\frac{1}{N}\sum_{n=1}^{N}
  \log q_\theta(\alv_n \mid \bx_n).
  \label{eq:npe}
\end{equation}
Once trained, inference on a new observation $\bx_0$ requires only a forward pass: samples are drawn directly from $q_\theta(\alv \mid \bx_0)$ without any further simulator calls.

\section{Architecture \& Training}
\label{sec:arch}

\begin{figure}[h!]
  \centering
  \includegraphics[width=\linewidth]{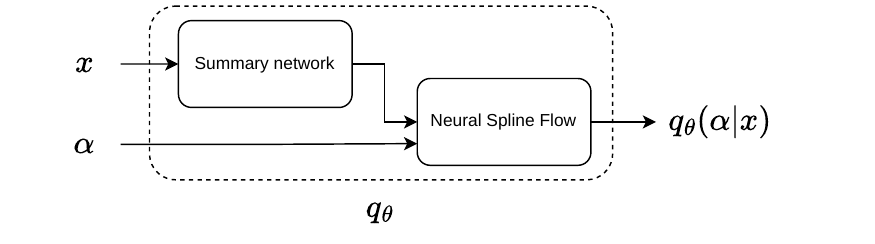}
  \caption{Summary of the neural architecture.}
  \label{fig:q}
\end{figure}

The neural network $q_\theta$ is built from a neural spline flow described in \cref{sec:de} and a summary network described in \cref{sec:sn}, which interact as described in \cref{fig:q}.

\subsection{Density Estimator}
\label{sec:de}

The conditional density $q_\theta$ is parameterized as a Neural Spline Flow~\cite{durkanNeuralSplineFlows2019a} (8~transforms, 256~hidden features) conditioned on a summary embedding of~$\bx$; both are trained jointly end-to-end by minimizing \cref{eq:npe}, using the \texttt{sbi} toolkit \cite{tejero-canteroSbiToolkitSimulationbased2020}.

\subsection{Summary Network}
\label{sec:sn}

Because the impulse response is a very large time series, the flow cannot easily condition on it directly: a summary network is first used to compress it into a lower-dimensional embedding.

The summary network is a CNN-based architecture that generates an embedding of the impulse response, which is then used to condition the normalizing flow.

\begin{figure}[ht]
  \centering
  \includegraphics[width=0.5\columnwidth]{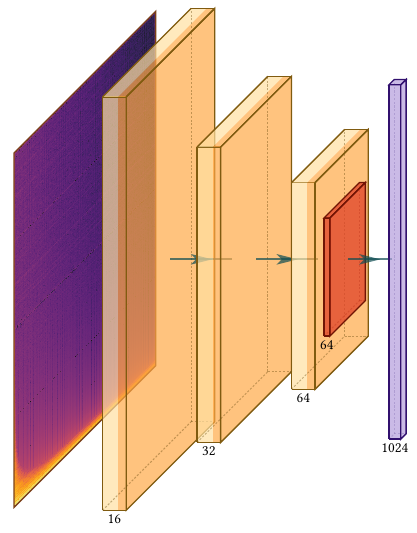}
  \caption{Architecture of one CNN branch.}
  \label{fig:cnn}
\end{figure}

The impulse response is first used to generate $3$~log-magnitude spectrograms at different resolutions (window sizes of $512$, $2048$, and $8192$).
Each of them is processed by a three-layer 2D CNN using $3{\times}3$ convolutions, followed by adaptive average pooling to a fixed $4{\times}4{\times}64$ spatial map, yielding a $1024$-dimensional vector per branch, as shown in~\cref{fig:cnn}.

The outputs of all branches are concatenated, producing a 3072-dimensional embedding. We then append the log-RMS amplitude of the raw IR to improve the reconstruction of $\mu$, as the IR's amplitude scale is inversely proportional to it.

The resulting $3073$-dimensional vector is finally projected via a two-layer MLP into a $256$-dimensional embedding, which is then passed on to the neural spline flow.

\subsection{Training}

The model is trained on 60\,000 simulator-generated pairs drawn from the prior by uniform sampling of the raw plate parameters, for 145~epochs with batch size~128, using the Adam optimizer with learning rate $10^{-3}$.
Training takes approximately 26~hours on an NVIDIA A100.
Early stopping on a held-out validation split is used to prevent overfitting.

\section{Sequential Refinement}
\label{sec:refine}

From the previously described architecture and training, we have a neural network $q_\theta$ that provides a good estimate for any plate but has to cover the entire parameter space at once. We would prefer a network that can be worse globally but has very good performance around the test observations we are specifically interested in.

To achieve this, for each test IR, we run a few extra rounds of simulation targeted at that specific observation and fine-tune the network on the results. This creates estimators that perform worse globally but better around the specific target point. This sequential refinement strategy is a standard technique in
SBI~\cite{cranmerFrontierSimulationbasedInference2020,tejero-canteroSbiToolkitSimulationbased2020}, and we follow the SNPE-C objective~\cite{greenbergAutomaticPosteriorTransformation2019}, which ensures the refined posterior remains correctly calibrated even though the new simulations are no longer drawn from the prior. Unlike offline training, this correction requires evaluating the prior density $p(\alv)$, which we compute in closed form, the main difficulty being the non-uniform distribution induced on the derived invariants.

\begin{figure}[ht]
  \centering
  \includegraphics[width=\columnwidth]{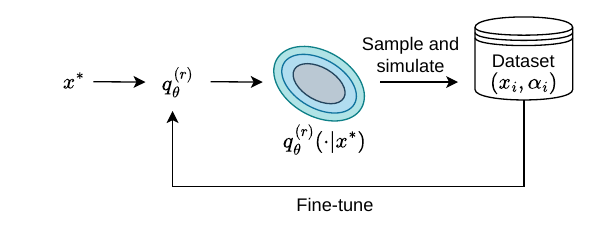}
  \caption{Sequential SBI refinement.}
  \label{fig:multiround}
\end{figure}

\needspace{2\baselineskip}
For test point $\bx^*$, for each round $r$:
\begin{itemize}
  \item \textbf{Propose.} Draw a batch of proposals $\alv_1,\dots,\alv_B$ from the current estimate $q_\theta^{(r-1)}(\alv \mid \bx^*)$, with $q_\theta^{(0)}$ being the initially trained density estimator from~\cref{sec:arch}.
  \item \textbf{Simulate.} Run the forward simulator at each proposal to obtain the corresponding impulse responses $\bx_1,\dots,\bx_B$, producing a dataset concentrated near the current estimate's mode. Each proposal is mapped to raw plate parameters by drawing $\rho$ uniformly at random and recovering $h$, $E$ and $T_0$ from the invariants.

  \item \textbf{Fine-tune.} Fine-tune $q_\theta$ using the atomic SNPE-C objective~\cite{greenbergAutomaticPosteriorTransformation2019}, which corrects for the non-prior proposal distribution to ensure the refined model still targets the true posterior $p(\alv \mid \bx^*)$.
\end{itemize}

For each test point, we fine-tune the model for 3~rounds, each round generating an extra 10\,000 IRs near the current estimate. For each test point, the $3$ rounds take a total of ${\sim}2.5$\,h to run on an A100 GPU.

The final point estimate is obtained by averaging over samples drawn from the refined density.
As we have found that refinement can cause the model to degenerate on some observations, we additionally guard against such failures with a selection step, described and evaluated in \cref{sec:results}.

\section{Results and Discussion}
\label{sec:results}

\begin{figure}[t]
  \centering
  \includegraphics[width=\columnwidth]{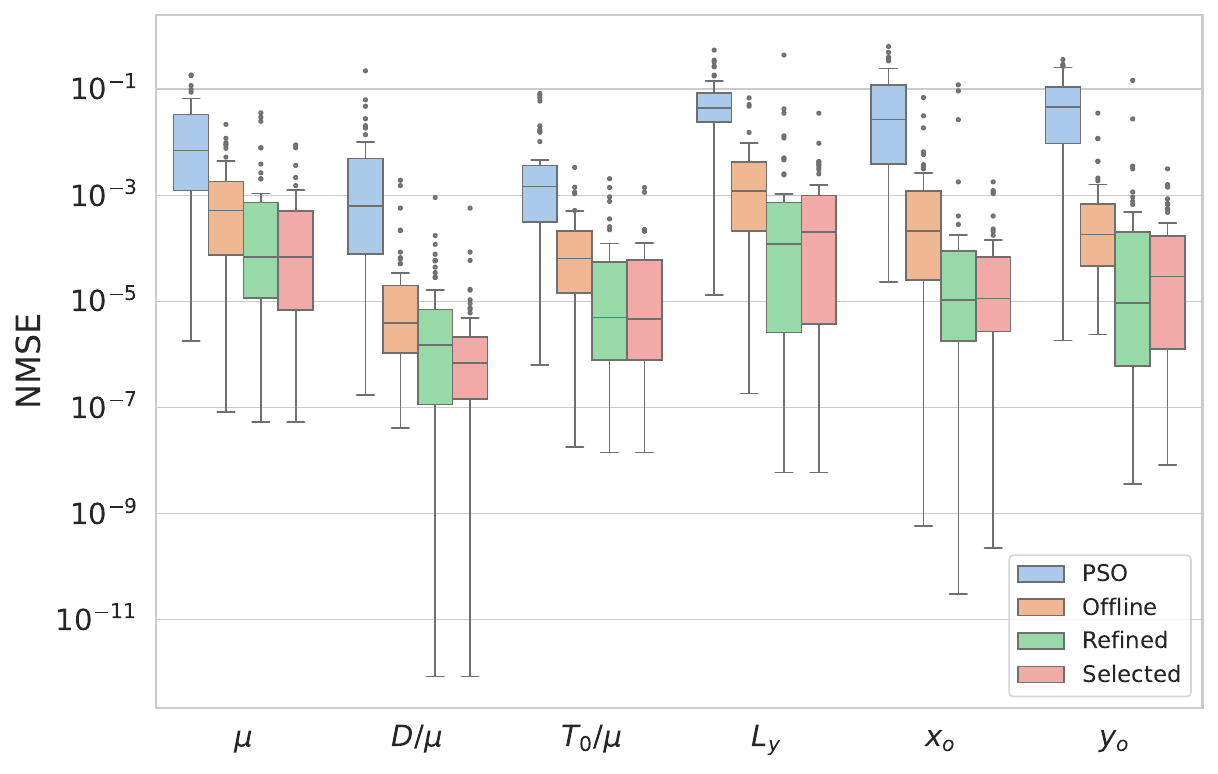}
  \caption{Distribution of the per-parameter NMSE across the $50$-IR test set.}
  \label{fig:nmse_per_param}
\end{figure}

To evaluate our model's performance, we generated a custom test dataset of $50$ random IRs following Task A's format.
\cref{tab:results} reports the per-parameter NMSE for different estimators, while \cref{fig:nmse_per_param} shows the corresponding error distributions.
The table reports averages, which on this task are dominated by a few catastrophic failures of the refined estimator; the figure shows what happens on typical IRs. Both views are needed to understand the results.

We compare three estimators. The \emph{offline} model is the base density estimator $q_\theta^{(0)}$ of \cref{sec:arch}, trained entirely offline and requiring no fine-tuning at inference time.
The \emph{refined} estimator is the per-observation model obtained directly by the sequential procedure of \cref{sec:refine}.
Finally, the \emph{selected} estimator is the one we submitted to the challenge\footnote{The version actually scored by the challenge used a uniform approximation of $p(\alv)$ in the SNPE-C correction (\cref{sec:refine}) rather than the exact density, reaching a slightly worse global NMSE of $0.94\times10^{-3}$ for the selected model (oracle: $0.45\times10^{-3}$) on the same test set.}: it tries to pick the best between the offline and refined estimators for each test IR, by keeping whichever yields the lower $\ell_2$ waveform reconstruction error.²²²

The \emph{oracle} represents the upper bound obtained by always picking the lower-NMSE estimate between the offline and refined methods for each IR.

\begin{table}[h!]
  \centering
  \caption{Average per-parameter NMSE ($\times 10^{-3}$, lower is better) on the $50$-IR test set. \emph{PSO} is the challenge's particle-swarm baseline.}
  \begin{minipage}{\columnwidth}
    \centering
    \resizebox{\columnwidth}{!}{%
      \begin{tabular}{l c @{\hskip 1.5em} c c c c c c}
        \toprule
        & & \multicolumn{6}{c}{Per parameter} \\
        \cmidrule(l){3-8}
        & Global & $\mu$ & $D/\mu$ & $T_0/\mu$ & $L_y$ & $x_o$ & $y_o$ \\
        \midrule
        PSO baseline & 50.56 & 26.16 & 9.72 & 8.46 & 80.43 & 95.18 & 83.39 \\
        \cmidrule(lr){1-8}
        Offline & 2.17 & 2.18 & 0.10 & 0.23 & 5.50 & 3.23 & 1.75 \\
        Refined & 3.80 & 2.49 & 0.03 & 0.14 & 11.45 & 4.93 & 3.74 \\
        Selected & 0.47 & 0.68 & 0.02 & 0.08 & 1.66 & 0.15 & 0.25 \\
        \cmidrule(lr){1-8}
        \textit{Oracle} & \textit{0.34} & \textit{0.44} & \textit{0.03} & \textit{0.08} & \textit{1.25} & \textit{0.10} & \textit{0.15} \\
        \bottomrule
    \end{tabular}}

    \label{tab:results}
  \end{minipage}
\end{table}

All our estimators improve substantially over the challenge's PSO baseline, whose global NMSE of $50.56\times10^{-3}$ remains close to the constant-predictor reference of $83\times10^{-3}$ (\cref{sec:setup}). The baseline particularly struggles to localize the readout position ($x_o$, $y_o$), which our models recover well.

In \cref{tab:results}, refinement appears to degrade performance: the refined average NMSE is worse than the offline one. However, looking at \cref{fig:nmse_per_param}, we can see this is an artifact of averaging. Refinement improves $40$ of the $50$ test IRs and cuts the median global NMSE from $0.75\times10^{-3}$ to $0.13\times10^{-3}$. But on the few IRs where it degenerates, it increases the error by several orders of magnitude. These failures, appearing as the long upper tails in \cref{fig:nmse_per_param}, are rare yet large enough to dominate the mean.

The selection step exists to catch exactly these failures and reduce the variance of refining: in \cref{fig:nmse_per_param}, the refined and selected medians are close for every parameter, so on most IRs, selection changes nothing. But discarding the few catastrophic cases is enough to cut the mean global NMSE by more than a factor of $4$ relative to the offline estimator, from $2.17\times10^{-3}$ to $0.47\times10^{-3}$, improving every individual parameter (\cref{tab:results}). The remaining gap to the oracle ($0.34\times10^{-3}$) measures how often the $\ell_2$ criterion picks the wrong estimate: finding a better criterion would hopefully close this gap.

Finally, this accuracy has a computational cost: refinement takes ${\sim}2.5$\,h per observation at inference time, whereas the offline estimator runs in less than a second.

\subsection{Posterior uncertainty and calibration}
\label{sec:posterior}

While the challenge forces us to produce a point estimate, one of the advantages of SBI is that we can access a full predicted distribution, which can notably help assess uncertainty.

NMSE says nothing about whether the posterior's uncertainty is trustworthy: an overconfident model can be accurate on average yet report completely wrong distributions. To verify the model's calibration, we use TARP~\cite{lemos2023sampling}. TARP estimates the posterior's expected coverage: across many held-out plates, it measures how often the true parameters fall within the posterior's $\alpha$-probability region, at each level $\alpha$. A well-calibrated posterior makes that fraction equal $\alpha$ at each level.

\begin{figure}[h]
  \centering
  \includegraphics[width=0.85\columnwidth]{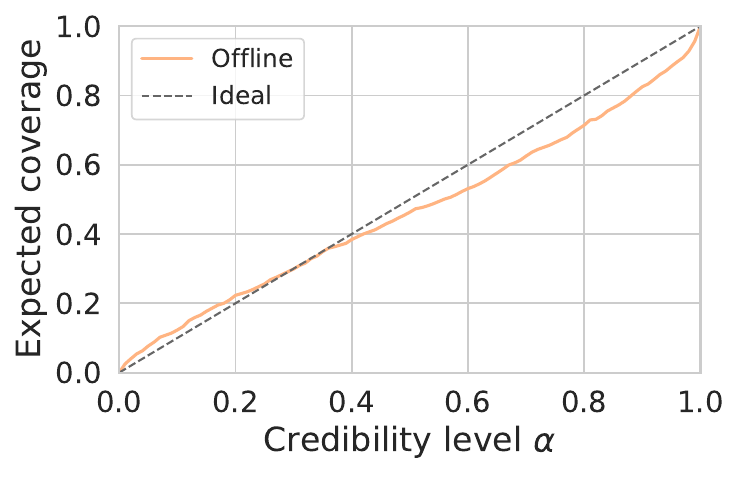}
  \caption{TARP expected coverage of the offline posterior over 500 held-out plates.}
  \label{fig:tarp}
\end{figure}

As shown in \cref{fig:tarp}, the offline posterior's coverage stays relatively close to the ideal diagonal, sitting slightly below it: our predicted distributions are a bit too narrow and mildly underestimate the long tails of the true posterior.


\section{Conclusion}

We cast plate-reverb system identification as simulation-based inference: a neural spline flow, conditioned on a multi-resolution CNN summary of the impulse response, yields a posterior over the six identifiable plate parameters. Finally, a per-observation refinement stage specializes it around each test IR, with a waveform-reconstruction heuristic falling back to the offline estimator when refinement degrades the fit. The submitted estimator reduces the global NMSE from $50.6\times10^{-3}$ for the challenge's PSO baseline to $0.47\times10^{-3}$.

The remaining gap to the per-IR oracle ($0.34\times10^{-3}$) makes a better criterion for accepting refined estimates the clearest lever for future work, alongside reducing the ${\sim}2.5$\,h per-observation cost of refinement, which currently stands against sub-second offline inference.

\section{Acknowledgements}
Many thanks to Sébastien Velut for his guidance on SBI methods.

This project was provided with computing and storage resources by GENCI at IDRIS under grant 2025-AD011017041 on the supercomputer Jean Zay's A100 partition.

\newpage

\bibliographystyle{IEEEtranDAFx}
\bibliography{dafx_references.bib} 

\end{document}